\documentclass[12pt]{article}
\usepackage{graphicx,a4}   
\usepackage{amssymb}        
\def\math{\mathsurround 0pt}
\def\oversim#1#2{\lower.5pt\vbox{\baselineskip0pt \lineskip-.5pt
        \ialign{$\math#1\hfil##\hfil$\crcr#2\crcr{\scriptstyle\sim}\crcr}}}

\begin{document}
\newcommand{\preprintno}[1]
{{\normalsize\begin{flushright}#1\end{flushright}}}

\title{\preprintno{{\bf ULB-TH/01-38}\\ hep-ph/0111435}Large extra dimensions, the galaxy power spectrum and the end of inflation}
\author{Malcolm Fairbairn\thanks{E-mail address: mfairbai@ulb.ac.be}\\
{\em Service de Physique Th\'eorique, CP225}\\
{\em Universit\'e Libre de Bruxelles, B-1050 Brussels, Belgium}\\
\\
Louise M. Griffiths\thanks{E-mail address: lmg@astro.ox.ac.uk}\, \thanks{Present address: Astrophysics, UNSW, Sydney, NSW 2052, Australia}\\
{\em Astrophysics, University of Oxford}\\
{\em Denys Wilkinson Building, Keble Road, Oxford OX1 3RH, England}}

\date{29th November 2001}

\maketitle
\begin{abstract}
We consider the production of gravitational KK modes via cosmological photon-photon and electron-positron annihilation in models with large factorisable extra dimensions.  We place constraints on this production using recent results from a joint analysis of the power spectra of the 2dF Galaxy Redshift Survey (2dFGS) and the cosmic microwave background (CMB) anisotropies. We obtain a more accurate upper limit for the temperature corresponding to matter-radiation equality and show that, even for the case of 6 extra dimensions and a fundamental scale of $1$ TeV, a period of inflation is required that ends at a temperature much lower than that of the QCD phase transition.

\end{abstract}

\section{Introduction}
One of the most fascinating possibilities which have come to light over recent years is that it is possible for there to be extra dimensions which are macroscopically large ($\sim 1$mm), provided only gravity is free to propagate in those directions.  The possibility is doubly interesting as it could cure the hierarchy problem, namely the huge gap between the Planck scale $M_{P}\sim 10^{19}$ GeV associated with the apparent strength of the gravitational interactions and the $\sim$ TeV scale of standard model particle physics.  If the standard model particles are confined to a brane whilst gravity is free to propagate in the full space-time, the extra dimensions effectively dilute the strength of gravity so that the observed value of Newton's gravitational constant in four dimensions is given by $\cite{arkani}$
\begin{equation}
4\pi G=\frac{4\pi}{M_{P}^2}=\left(\frac{1}{RM_{F}}\right)^{d}\frac{1}{M_{F}^{2}}
\label{bigg}
\end{equation}
where $R$ is the radius of the extra dimensions, $d$ is the number of large extra dimensions, G is Newton's constant in 3+1 dimensions and $M_{F}$ is the fundamental scale of both gravitational and gauge interactions in the $3+1+d$ dimensional theory.

Because the extra dimensions are topologically compact, the possible momenta of gravitational waves which propagate in the extra dimensions are subject to periodic boundary conditions.  The momenta of these Kaluza-Klein (KK) modes must therefore have values which are integer multiples of that of the longest wavelength gravitational wave which can fit around the extra dimension.  Since we perceive momentum in the extra dimension as mass, these KK modes look to us like a spectrum of massive particles with an infinite ladder of possible masses.  The gravitational KK modes are excited like normal gravitational waves, by coupling to the stress-energy of the standard model particles on the brane.  

The more traditional view of higher dimensions saw the extra dimensions compactified to a scale comparable with the Planck length.  In this situation the mass of the KK modes, whose characteristic mass scale is set by the inverse of the compactification scale, is so high that one would never expect them to be excited (although see $\cite{kolbslan}$).  In the large extra dimensional case, the mass of the KK modes can typically be much smaller than many energies which occur in astrophysical and cosmological situations.  Since the KK modes are effectively just gravitational waves, their coupling to the stress-energy of the rest of the universe is Planck suppressed, and it is still difficult to excite them even from a plasma with a characteristic energy scale many orders of magnitude higher than their mass.  However, the very large number of kinematically available modes below the energies experienced in supernovae or in the early universe mean that KK mode generation can be significant.

The neutrinos detected from supernova 1987a in the Large Magellanic Cloud successfully accounted for the majority of the gravitational energy expected to be released during the core collapse of the precursor $\cite{raffelt}$.  Consequently a constraint can be made on the energy carried away by the generation of KK modes during the collapse $\cite{sn,hanhart}$.  A stronger constraint may be obtained by considering the subsequent decays of the KK modes.  Depending on their mass, the KK modes typically have a lifetime comparable with the Hubble time or longer $\cite{lykken}$ and their decay would contribute to the cosmic gamma ray background $\cite{hall,decay,hannestad}$.  In this way constraints can be placed on the production of KK modes in supernovae and also in the early universe.  This method cannot easily constrain the case of 6 extra dimensions, since in this situation, KK modes with higher masses are produced which decay more quickly, reducing their impact on the cosmic gamma-ray background.

The size of extra dimensions is also constrained by the requirement of successful baryogenesis, but this constraint is model dependent $\cite{baryo}$.

Another way to constrain the number and size of extra dimensions is to consider their effect on the epoch of matter-radiation equality, since KK modes redshift as matter as opposed to radiation.
In $\cite{fairbairn}$, one of us tried to show that production of KK modes in the early universe due to nucleon-nucleon gravi-bremsstrahlung could instigate an early epoch of matter-radiation equality resulting in too short an age for the universe.  The purpose of this paper is to act upon two suggestions which have been made since then and therefore update these constraints.  The first is the observation of Hannestad $\cite{hannestad}$ that photon-photon and electron-positron annihilation will lead to more KK mode production in the early universe than nucleon-nucleon gravi-bremsstrahlung, the dominant mechanism in supernova cores.  The second suggestion $\cite{barrow}$ is that we can tie down the epoch of matter-radiation equality more strongly by considering the turn-over in the matter power spectrum which occurs at the angle on the sky corresponding to the comoving scale that enters the horizon at equality.  We are therefore able to more accurately constrain the production of KK modes, not by considering the age of the universe but by using the results of a combined likelihood analysis of the power spectra of the 2dF Galaxy Redshift Survey (2dFGRS) and the cosmic microwave background (CMB) anisotropies $\cite{efst}$.

If the extra dimensions are large enough to cure the hierarchy problem, we find that a period of inflation must exist which ends at a very low energy ($\sim$10-100 MeV) since many KK modes are produced in the cosmological plasma.  Inflation cannot occur at energies less than $\sim$ 1 MeV as this would compromise the successful predictions of primordial nucleosynthesis.  However, it has been shown that inflation can end extremely close to the beginning of nucleosynthesis $\cite{kawasaki}$. 

We will go over the Friedman equations and write down the expressions for emissivity due to photon-photon and electron-positron annihilation.  Then for each value of the fundamental scale $M_F$ in $3+1+d$ dimensions we will calculate how low the end of inflation has to be in order to prevent KK modes inducing early matter-radiation equality.  Finally we will discuss these results in the context of the possibility of a low temperature for the end of inflation.

\section{Cosmology}
If one wishes to neglect the presence of energy in the bulk on the cosmological evolution, one needs to ensure that the total energy in the bulk is much less than that on the brane.  This is because the brane is in causal contact with all the stress energy from the bulk soon after the extra dimension has become stabilised.  Assuming the brane is of width $M_F^{-1}$ in the extra dimensions, the condition for normal cosmological evolution becomes
\begin{equation}
\rho_{brane}^{4+d} V_{brane}>\rho_{bulk}^{4+d} V_{bulk}
\end{equation}
which from equation ($\ref{bigg}$) gives us
\begin{equation}
\rho_{brane}^{4+d}>\left(\frac{M_{P}}{M_{F}}\right)^{2}\rho_{bulk}^{4+d}.
\end{equation}
Consequently we need to ensure that the energy density in the bulk is less than that on the brane by at least a factor $M_{F}^{2}/M_{Pl}^{2}\ll 1$ in order to ensure the bulk energy density is not affecting the cosmological dynamics $\cite{linde}$.  This never occurs in the work which follows, although we have to assume the bulk is free of energy density to start with.  One would expect this after a period of inflation on the brane, and this is one of the assumptions we make here.

Energy conservation for the KK modes in an expanding universe requires
\begin{equation}
\frac{d\rho_{KK}}{dt}=\frac{d\epsilon_{KK}}{dt}-3H\rho_{KK}
\end{equation}
where $\rho_{KK}$ is the four dimensional effective density of the KK modes, H is Hubble's constant and $\epsilon_{KK}$ is the emissivity of the plasma into KK modes.  Since we feel the gravitational effect of everything in the bulk, the effective energy density in KK modes is equal to the amount of energy density lost from the brane in their emission. 

The expression for time as a function of temperature in the radiation dominated era is
\begin{equation}
t=\frac{0.301}{\sqrt{g_{*}G}T^{2}}; \qquad H=1.66\sqrt{g_{*}G}T^{2}
\label{time}
\end{equation}
where $g_{*}$ is the total number of relativistic degrees of freedom in the plasma which are given in the appendix. Then we can write
\begin{equation}
\frac{d\rho_{KK}}{dT}=-\frac{0.602}{\sqrt{g_{*}G}T^3}\frac{d\epsilon_{KK}}{dt}+\frac{3\rho_{KK}}{T}
\label{wrtemp}
\end{equation}
which admits analytic solutions if the emissivity is a power law with respect to temperature

The dominant mechanisms responsible for the production of massive KK excitations in the early universe will be photon-photon and electron-positron annihilation.  At temperatures far above the electron mass the expressions for KK mode emission for these two processes do not contain a Boltzmann factor, so if we neglect the less important nucleon-nucleon gravi-bremsstrahlung, we can obtain analytic solutions.  The relevant emissivity rates were presented in $\cite{Barger:1999jf}$.  

Once the spin averaged total cross section for the process $\gamma\gamma\rightarrow$KK is obtained from the Feynman diagrams of $\cite{lykken}$, the emissivity into KK modes is given by integration over two Bose-Einstein ensembles of photons. After carrying out these integrals and the integrals over the KK modes which make up the phase space one is left with the volume averaged emissivity $\cite{Barger:1999jf}$
\begin{equation}
\frac{d\epsilon_{\gamma\gamma\rightarrow KK}}{dt}=\frac{2^{d+3}\Gamma(\textstyle{\frac{d}{2}}+3)\Gamma(\textstyle{\frac{d}{2}}+4)\zeta(\textstyle{\frac{d}{2}}+3)\zeta(\textstyle{\frac{d}{2}}+4)}{(d+4)\pi^{2}}\frac{T^{d+7}}{M_{F}^{d+2}}={\mathcal{C}}_{\gamma\gamma}\frac{T^{d+7}}{M_{F}^{d+2}}
\label{gamemiss}
\end{equation}
where $d$ is the number of extra dimensions, $\Gamma$ is the normal Gamma function and $\zeta$ is the Riemann zeta function.

In a similar way, the $e^{+}e^{-}\rightarrow KK$ volume emissivity is obtained by integrating the cross section over two Fermi-Dirac distributions
\begin{eqnarray}
\frac{d\epsilon_{e^{+}e^{-}\rightarrow KK}}{dt}=\frac{2^{d}I(d)}{(d+4)\pi^{2}}\frac{T^{d+7}}{M_{F}^{d+2}}={\mathcal{C}}_{e^{+}e^{-}}\frac{T^{d+7}}{M_{F}^{d+2}}
\label{eeemiss}
\end{eqnarray}
where the integral factor $I(d)$ is given by
\begin{equation}
I(d)=\int_{0}^{\infty}dx\int_{0}^{\infty}dy\frac{(xy)^{\textstyle{\frac{d}{2}}+2}(x+y)}{[exp(x-\textstyle{\frac{\mu_{e}}{T}})+1][exp(y+\textstyle{\frac{\mu_{e}}{T}})+1]}.
\label{integral}
\end{equation}

Here $\mu_{e}$ is the chemical potential for the electron.  In the early universe at temperatures much higher than the electron mass, the electron distribution will be in thermal equilibrium with the surrounding radiation bath.  The Helmholtz free energy $F=E-TS$ is minimised in equilibrium and as the number of particles $N$ in a system evolves to equilibrium, one would expect that a small fluctuation in $N$ would have little effect on the energy of the system $\cite{peacock}$.  The first law for the free energy is given by
\begin{equation}
dF=-SdT-PdV+\mu dN
\end{equation}
so in the thermal bath of the early universe, $dF/dN=0\rightarrow\mu=0$ which allows one to find analytic solutions for equation ($\ref{integral}$) which are given in the appendix.

Below 4.3 MeV we will neglect any further contribution from the electron-positron annihilation and consider only KK modes produced from photon-photon annihilation.  This approximation will slightly underestimate the KK mode generation and therefore the strength of our conclusions.

Combining ($\ref{gamemiss}$) and ($\ref{eeemiss}$) we are left with a total emissivity for photon-photon and electron-positron annihilation of the form
\begin{equation}
\frac{d\epsilon}{dt}=\left({\mathcal{C}}_{e^{+}e^{-}}+{\mathcal{C}}_{\gamma\gamma}\right)\frac{T^{d+7}}{M_{F}^{d+2}}
\label{produce}
\end{equation}
where the ${\mathcal{C}}$s are set by the expressions for the emissivity presented above.  Equation ($\ref{wrtemp}$) is now
\begin{equation}
\frac{d\rho_{KK}}{dT}=-\frac{0.602}{\sqrt{g_{*}G}}\left({\mathcal{C}}_{e^{+}e^{-}}+{\mathcal{C}}_{\gamma\gamma}\right)\frac{T^{d+4}}{M_{F}^{d+2}}+\frac{3\rho_{KK}}{T}.
\label{diff}
\end{equation}
If we define some temperature $T_0$ where the density is given by $\rho_{KK0}$ we then find the general solution
\begin{equation}
\rho_{KK}(T)=\frac{0.602}{\sqrt{g_{*}G}}\frac{{\mathcal{C}}_{e^{+}e^{-}}+{\mathcal{C}}_{\gamma\gamma}}{(2+d)M_F^{2+d}}\left(T^3 T_0^{2+d}-T^{5+d}\right)+\left(\frac{T}{T_0}\right)^3 \rho_{KK0}
\label{sol}
\end{equation}
which we can solve in between each of the freeze-out/annihilation temperatures using the degrees of freedom listed in the appendix.

\section{The Epoch of Matter-Radiation Equality}

The turn-over in the matter power spectrum occurs at a characteristic scale corresponding to the comoving size of the horizon at the epoch of matter-radiation equality.  Therefore, the epoch of matter-radiation equality can be strongly constrained by the matter power spectrum extracted from large scale structure surveys.  The overall shape of the cold dark matter (CDM) transfer function, and thereby the turn-over in the matter power spectrum, is defined by the parameter combination $\Omega_m h$ (where $\Omega_m$ is the non-relativistic matter density parameter and $h$ is the Hubble constant $H_0$ in units of 100kms$^{-1}$Mpc$^{-1}$).  Limits on $\Omega_m h$ together with estimates of $h$ enable us to give tight constraints on the temperature at the epoch of matter-radiation equality, $T_{EQ}$.

The 2dFGRS is one of the most advanced large scale structure surveys to date, designed to measure redshifts for approximately 250 000 galaxies $\cite{colless}$.  Efstathiou et al. $\cite{efst}$ perform a combined likelihood analysis of the power spectra of the 2dFGRS and the CMB anisotropies under the assumptions that the galaxy power spectrum on large scales ($k < 0.15 h$ Mpc$^{-1}$) is directly proportional to the linear matter power spectrum and that the initial fluctuations were adiabatic, Gaussian and well described by power laws with scalar and tensor indices of $n_s$ and $n_t$.  Adding the constraints on the baryon content of the universe from big-bang nucleosynthesis (BBN) $\cite{burles}$, they give a range for $\Omega_m h$ of
\begin{equation}
0.16 < \Omega_m h < 0.21 \qquad (95\% \, {\rm{C.L.}}\,, {\rm{best}} \, {\rm{fit}} = 0.19)\,.
\end{equation}

Since the density of non-relativistic matter scales as 1/$a^3$ (where $a$ is the scale factor) and that of relativistic matter scales as 1/$a^4$ and given that temperature is inversely proportional to $a$,
\begin{equation}
T_{EQ}/T_0 = a_0/a_{EQ} = \Omega_m/\Omega_{\rm rel} \,,
\end{equation}
where $\Omega_m$ and $\Omega_{\rm rel}$ are the present values for the non-relativistic and relativistic matter density parameters respectively.  Taking $T_0 = 2.725 \pm 0.002$ K $\cite{mather}$, $\Omega_{\rm rel} h^2$ is evaluated to be $4.15 \times 10^{-5}$, assuming that the relativistic degrees of freedom are the same at BBN and CMB epochs as the simplest models predict.  Combining the constraints on $\Omega_m h$ with the recent HST Hubble key project result $\cite{freedman}$
\begin{equation}
h = 0.72 \pm 0.16 \qquad (95\% \, {\rm{C.L.}}),
\end{equation}
we infer an upper limit for $T_{EQ}$ of
\begin{equation}
T_{EQ} < 11 \times 10^3 K \qquad (95\% \, {\rm{C.L.}}) \,.
\end{equation}

The constraints on the baryon content of the universe from primordial nucleosynthesis suggest that $\cite{burles}$
\begin{equation}
\Omega_b h^2=0.02 \pm 0.002 \qquad (95\% \, {\rm{C.L.}}) \,,
\end{equation}
so the ratio between the baryonic and total matter in the universe is
\begin{equation}
\frac{\Omega_b h^2}{\Omega_m h^2}=0.146 \pm 0.039 \qquad (95\% \, {\rm{C.L.}}).
\end{equation}
This tells us that the density of the KK modes at $T_{EQ}$ must be less than or equal to $89\%$ of the radiation density given by
\begin{equation}
\rho_{\gamma}=\frac{g_{*}\pi^{2}}{30}T^{4}
\end{equation}
so the density of KK modes at equality can at most be
\begin{equation}
\rho_{KK}(T_{EQ})\le 0.89\frac{g_* \pi^2 T_{EQ}^4}{30}=7.9\times 10^{-25} MeV^4.
\label{con}
\end{equation}
Using equation ($\ref{sol}$) we can then calculate the maximum temperature at which inflation has to end in order for the production of KK modes due to the two annihilation processes creating so much matter energy density in the early universe that the bound ($\ref{con}$) cannot be met.  First we need to quantify how many of the KK modes decay back into particles on the brane before the epoch of matter radiation equality.

\section{Decay of KK modes}

A number of the more massive KK modes decay into standard model particles on the brane before matter-radiation equality and this needs to be taken into account in our constraints.

The decay of KK modes into standard model particles on the brane has been calculated in $\cite{lykken}$.  For KK mode masses well below that of the Higgs boson or the intermediate vector bosons, the bosonic decays are given by
\begin{equation}
\Gamma_{\gamma\gamma + gg}(m_{KK})=9\frac{m_{KK}^3}{10M_{Pl}^2}
\end{equation}
where the 9 comes from the fact there are 8 gluons and 1 photon.  Decay into gluons only occurs when the KK modes have a larger mass than the lower lying hadrons.  The decay into fermions is 
\begin{equation}
\Gamma_{\bar{f}f}(m_{KK})=N_c\frac{m_{KK}^3}{20M_{Pl}^2}\left(1-4\frac{m_f^2}{m_{KK}^2}\right)^{3/2}\left(1+\frac{8}{3}\frac{m_f^2}{m_{KK}^2}\right).
\end{equation}
The decay width for all processes is therefore roughly proportional to $m_{KK}$ cubed and the decay lifetime of the KK mode is related to the various decay widths by
\begin{equation}
\tau(m_{KK})=\frac{1}{\sum_n\Gamma_n(m_{KK})}.
\end{equation}

\begin{figure}[tb]
\centering
\includegraphics[width=10.5cm,height=8cm]{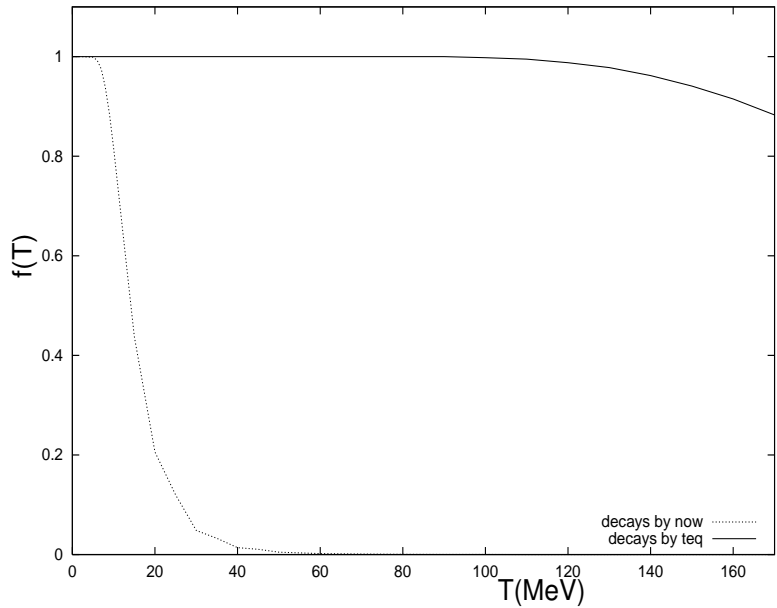}
\caption{\label{f}Fraction f(T) of KK modes with decay lifetime $\tau(m_{KK})$ greater than $t_{EQ}$ and $t_{now}$}
\end{figure}

Since we are not concerned about the flux of decay products at any one time we will assume that all KK modes of mass $m_{KK}$ decay together at a time $t=\tau(m_{KK})$ after their creation.  The mass of the KK mode with decay lifetime $\tau$ equal to $t_{EQ}\approx 4.7\times 10^{5}$ years is $\approx 1600$ MeV.  It was shown in $\cite{hall}$ that the creation of KK modes of a particular mass $m_{KK}$ is proportional to
\begin{equation}
\frac{dn(m_{KK})}{dt}+3Hn(m_{KK})\propto m_{KK}^5 T \mathcal{K}_1(\frac{m_{KK}}{T})
\end{equation}
where $n(m_{KK})$ is the number density of KK modes of mass $m_{KK}$,  H is the Hubble expansion and $\mathcal{K}_1$ is a Bessel function of the first kind.  Consequently the fraction $f(T)$ of the KK mode energy density comprising of KK modes with decay lifetimes $\tau$ longer than $t_{EQ}$ is given by
\begin{equation}  
f(T)=\frac{\int_0^{1600}m^6 T\mathcal{K}_1(\frac{m}{T}) dm}{\int_0^{\infty}m^6 T\mathcal{K}_1(\frac{m}{T}) dm}.
\end{equation}
The solution of this equation can be seen in figure $\ref{f}$

We can then use $f(T)$ to modify equation (\ref{diff}) which means that our solution (\ref{sol}) is only valid at low temperatures where $f(T)$ is effectively unity.  We therefore have to integrate (\ref{diff}) numerically at high temperatures. 

Figure $\ref{f}$ also shows the number of KK modes produced at different temperatures which have decayed by today and will therefore have a diminished effect upon the gamma-ray background.  Equation $\ref{produce}$ shows that the higher the number of extra dimensions, the more temperature sensitive the production of KK modes, so for the case of $d=6$ most of the modes produced will be those from the highest temperature which exists in the radiation dominated era.  Consequently, the majority of those modes will have decayed by today if the end of inflation occurs at some temperature above about 40 MeV.  Therefore although our constraints are not as strong as those of $\cite{decay}$ (if one is willing to entertain the possibility of inflation below the QCD phase transition)  we can start to constrain the parameter space of 5 and 6 large extra dimensions.

\section{Results and Discussion}

\begin{figure}[tb]
\centering
\includegraphics[width=10.5cm,height=8cm]{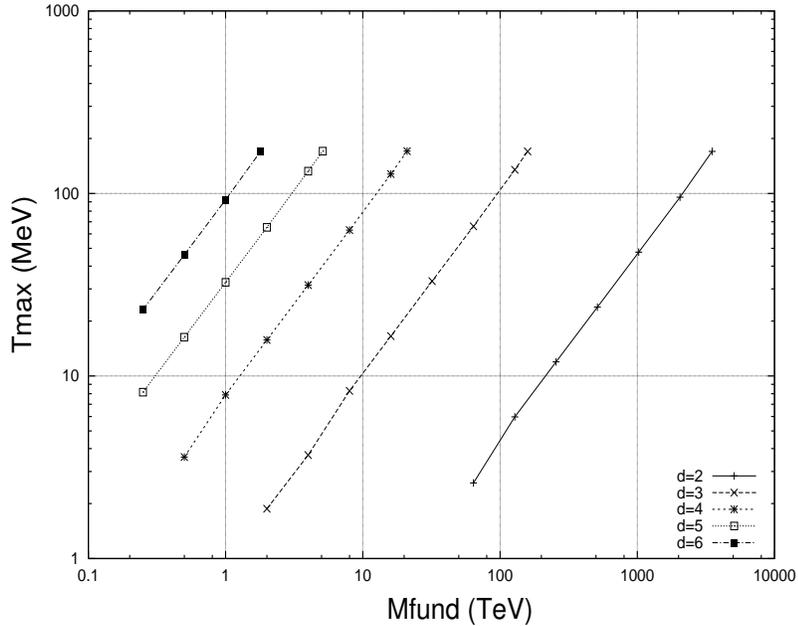}
\caption{\label{plots}Maximum temperature required for the end of inflation for various numbers $d$ of the extra dimensions and values of the fundamental scale}
\end{figure}

Given that $T_{EQ}$ must be less than $11\times 10^3 K$ we can proceed to ensure that KK modes are not over-produced such as to violate this limit.  For each value of $M_F$ and each number of extra dimensions $d$ one can calculate the energy scale corresponding to the end of inflation in order to prevent violation of equation $\ref{con}$.  These results are shown in figure $\ref{plots}$.

They show that in it is necessary to have a period of inflation below the QCD phase transition in order to use large extra dimensions to cure the hierarchy problem.  It is not possible to consider the effects of the production of KK modes before the QCD phase transition, as it is not clear whether or not a small amount of inflation may occur at that time (see $\cite{bonometto}$ and references within).  In table $\ref{end}$ we list the maximum temperature at which inflation must occur in order to prevent over production of KK modes from the plasma in the case of $M_F = 1$TeV.
\begin{table*}
\begin{center}
\caption[]{\label{end} Maximum temperature for inflation to end for different numbers of extra dimensions and $M_F$=1 TeV}
\begin{tabular}{|c|c|}\hline
d	&Tmax (MeV) \\ \hline
2	&forbidden  \\ \hline
3	&forbidden  \\ \hline	
4	&7.88 MeV   \\ \hline 
5	&32.6 MeV   \\ \hline		
6	&92.3 MeV   \\ \hline
\end{tabular}
\end{center}
\end{table*}
  Even for $d=6$ extra dimensions it is necessary to have a period of inflation ending at a temperature lower than $120$MeV to allow a fundamental scale as low as a TeV.  This is striking as in some sense the case of 6 isotropic extra dimensions is a very natural one in the context of a 10 dimensional string theory.   

Calculations of the time taken for neutrinos to thermalise after some epoch of re-heating, which could take place either as a results of the decay of short lived relics or at the end of inflation, show that it is possible for the re-heat temperature to be as low as 0.7 MeV without affecting the successful nucleosynthesis prediction for helium abundance $\cite{kawasaki}$.  If this is true, it is not possible to place very tight constraints on the size and number of extra dimensions using our method, and one must consider the effect of the subsequent decay of the KK modes $\cite{hannestad}$.  

We are left with the conclusion that even in the case of $6$ extra dimensions we cannot avoid the requirement of a period of inflation at a very low temperature (less than $T_{QCD}$) to dilute the KK modes down to a level compatible with the latest data from galaxy surveys.  

\section*{Acknowledgements}
Thanks to John Barrow, Nicolas Borghini, Ed Copeland, Steen Hannestad, Steve Moody, Liam O'Connell and Bernard Pagel for useful conversations and/or suggestions.  MF is funded by the FNRS and LMG by PPARC.  LMG is currently a Visiting Associate at the University of New South Wales and acknowledges the use of their facilities.
\section*{Appendix}

The values used for each of the freeze-out or annihilation temperatures were obtained from $\cite{peacock}$ and are listed in table $\ref{freeze}$.
\begin{table*}
\begin{center}
\caption[]{\label{freeze} Degrees of freedom of the plasma at different temperatures}
\begin{tabular}{|c|c|c|}\hline
Particle	&Freezeout/annihilation &$g_{*}$ after freezeout \\ \hline
$T_{QCD}$ (gluons)&170 MeV		&14.25  \\ \hline 
$\mu^{\pm}$ 	&109 MeV		&10.75	\\ \hline		
$\nu$		&$\approx$ 4.5 MeV	&6.86	\\ \hline
$e^{\pm}$ 	&4.3 MeV		&3.36   \\ \hline
\end{tabular}
\end{center}
\end{table*}
\begin{table*}
\begin{center}
\caption[]{\label{phase} Values of the phase space integral ($\ref{integral}$) in the limit of negligible chemical potential}
\begin{tabular}{|c|c|c|c|c|c|}\hline
$d$	&2	&3	&4	&5	&6	\\ \hline
$I(d)$	&265.14	&1147.2	&5518.5	&29225 	&169042	\\ \hline
\end{tabular}
\end{center}
\end{table*}

\end{document}